\documentclass{article}

\newcommand{\ba}{\begin{eqnarray}}
\newcommand{\ea}{\end{eqnarray}}

\def \CSthreebarone {\mbox{$\bar{\mathbf{3}}_{1/2}$}}
\newcommand{\SpSp}[2]{ \mbox{$\vec{\sigma }_{#1}.\vec{\sigma }_{#2}$}}
\newcommand{\lala}[2]{ \mbox{${\vec{\lambda}_{#1}\cdot
     \vec{\lambda}_{#2}}$}}

\begin{document}

\begin{center}
{\Large \textbf{The colour triplet $qq\bar{q}$ cluster\\ and pentaquark
models}}
\vspace{25mm}

{\large H. Hogaasen}\\
{\em Department of Physics
University of Oslo\\
Box 1048  NO-0316 Oslo Norway\\
hallstein.hogasen@fys.uio.no}\\

\vspace{10mm}

and\\

\vspace{10mm}

{\large P. Sorba} \\
{\em Laboratoire d'Annecy-le-Vieux de Physique Th\'{e}orique (LAPTH) \\
9 chemin de Bellevue, B.P. 110, F-74941 Annecy-le-Vieux Cedex,
FRANCE\\
sorba@lapp.in2p3.fr}

%\date{May 28., 2004}
%\maketitle

\end{center}

\vspace{20mm}

\begin{abstract}
We study the properties of the colour triplet $qq\bar{q}$ quark
cluster when flavour symmetry is broken. The relevance of such a
cluster for some models of pentaquarks is then examined in the
light of recent experimental signals.

\end{abstract}

\vfill \vfill

\rightline{LAPTH-1048/04}
\rightline{hep-ph/0406078}
\rightline{May 2004}

\clearpage
\pagestyle{plain}
\baselineskip=18pt

%\section{Introduction}

The interest in multiquark states has increased very much since
the reports of possible observations of
  exotic baryons have come.
Whereas recent predictions of these states where $\Theta^+(1540)$
\cite{thetaexp} ,\cite{Alt:2003vb} ,\cite{H1} is at a
surprisingly low mass, came \cite{dpp} from chiral soliton models,the studies
of baryons with more than three quarks go back more than a quarter
of a century. At that time one made models where "coloured ions"
were bound together by colour-electric flux tubes\cite{Chan:1978nk} .
The mass defect
due to colour-magnetism were mostly made in the flavour symmetric
limit where  group theoretical mass formula was applied,the mass
defect can then be expressed by the quadratic Casimir operators
for the SU(2)-spin,the SU(3)-colour and the SU(6)-colourspin group
\cite{jaffescalar},\cite{Hogaasen:1978jw}. In
the cases where colour-spin,colour and spin for quarks and
  antiquarks can be
simultaneously quantized together with the same operators
  for the whole system,the results are quite easily generalized
to flavour symmetry breaking. In other cases not. An example which
has aroused some interest lately is the case when we have two
quarks (labeled 1 and 2) and an antiquark (3), all in a relative
s-wave and coupled to spin 1/2 and colour 3. This is the type of
"triquark" states that has recently been used, together with a spin
zero diquark state carrying colour $\bar{3}$, to make pentaquark
states of spin ${1/2}^+$ when the triquark and diquark are
separated by a L=1 orbital angular momentum.

The modest purpose of this letter is two fold. First we present a
detailed and algebraically correct analysis of the colourmagnetic
 \cite{dgg} interaction Hamiltonian in case of complete flavour
symmetry breaking. Then we examine to what extent the model
constituted by two-colour triplets $(qq\overline{q})^c_s$ with
$c=3$ and $s=1/2$ and $(qq)^c_s$ with $c=\overline{3}$ and $s=0$,
separated by an $L=1$ relative angular momentum, is well adapted
or not suitable to describe the states $\Theta^+(1540)$ and
$\Xi^{--} (1862)$. As can be immediately noticed by elementary
group theory computations, states of the
$(qq\overline{q})^3_{1/2}$ are {\em mixtures} of states of colour
$SU(3)_c$, spin $SU(2)_s$ and colour spin $SU(6)_{cs}$
representations, and that implies some care in their treatment.

When all spatial degrees of freedom are integrated out we have an
interaction Hamiltonian over colour spin space which is the usual

\begin{equation}\label{1}
    H_{\mathrm{CM}} = - \sum_{i,j} C_{ij} \lala{i}{j} \SpSp{i}{j}
\end{equation}

Here the coefficients $C_{ij}$ are, among other things, dependent
on the quark masses and properties of the spatial wave functions
of the quarks and the antiquark in the system. The solution of the
eigenvalue problem of the Hamiltonian above is therefore of
interest, not only in spectroscopy, but in all reactions where an
antiquark interact with a system of two quarks. The two quarks
$q_1$  and $q_2$ can be coupled to colour  $\bar{\mathbf{3}}$ or
$\mathbf{6}$, to spin 0 or spin 1. Together with the antiquark
$\bar{q}_3$, spin and colour couplings are such that the cluster
carries total colour 3 and spin 1/2.

It follows that  the space on which the Hamiltonian (eq.\ref{1})
acts over is four dimensional and a natural basis is provided with
the four states\\

\begin{eqnarray}\label{2}
    \phi_1 &=& |(q_1 q_2)^6_1 \rangle \otimes |
    (\overline{q}_3)^{\overline{3}}_{1/2} \rangle \nonumber \\
    \phi_2 &=& |(q_1 q_2)^{\overline{3}}_1 \rangle \otimes |
    (\overline{q}_3)^{\overline{3}}_{1/2} \rangle \nonumber \\
        \phi_3 &=& |(q_1 q_2)^6_0 \rangle \otimes |
    (\overline{q}_3)^{\overline{3}}_{1/2} \rangle \nonumber \\
        \phi_4 &=& |(q_1 q_2)^{\overline{3}}_0 \rangle \otimes |
    (\overline{q}_3)^{\overline{3}}_{1/2} \rangle
\end{eqnarray}
The notation here is
\begin{equation}\label{3}
\phi_i = |(q_1 q_2)^c_s \rangle \otimes |
    (\overline{q}_3)^{\overline{3}}_{1/2} \rangle
\end{equation}\\
where $c$ is the colour, $s$ is the spin of the two quarks $q_1$
and $q_2$ . The coupling with the antiquark state  $\bar{q}_3$
$|3,\CSthreebarone \rangle $  is to a total
 colour triplet with spin one half. For completeness, let us recall
the following product decompositions of $SU(3)$ representations:
\begin{equation}\label{4}
    3 \times 3 = \overline{3} + 6 \ \ \ \ ; \ \ \ \ \overline{3}
    \times \overline{3} = 3 + \overline{6} \ \ \ \ ; \ \ \ \ 6
    \times \overline{3} = 3 + 15
\end{equation}
The states  $\phi_1$  and $\phi_4$  have  two quarks which are
coupled symmetrically in colour-spin and are therefore belonging
to the $(6 \times 6)_S = 21$ dimensional representation of
$SU(6)_{cs}$, the states $\phi_2$  and $\phi_3$ are antisymmetric
in colour-spin of the two quarks and fall in the $(6 \times 6)_A =
15$ dimensional representation.

Note that if the two quarks are identical in flavour, the states
$\phi_1$ and $\phi_4$ vanish due to the Pauli principle.

A way to explicitly compute the $4 \times 4$ matrix representing
$H_{CM}$ relative to the $(qq\overline{q})^3_{1/2}$ triplet is to
study separately the colour part and the spin part namely:
\begin{equation}\label{5}
    H_C = - \sum_{i,j} C_{ij} \overrightarrow{\lambda}_i
    \cdot \overrightarrow{\lambda}_j \ \ \ \ \ H_S = -  \sum_{i,j} C_{ij} \overrightarrow{\sigma}_i
    \cdot \overrightarrow{\sigma}_j
\end{equation}
and then to perform a kind of "tensor product" of the two
so-obtained $2 \times 2$ matrices.

Let us consider the colour-action part. Then, when acting by $H_C$
on the $\phi_i$'s, it will be convenient to express $| (q_1 q_2)^c
(\overline{q}_3)^{\overline{3}} \rangle^3$ where $c =6$ or
$\overline{3}$ in terms of  $| (q_1 \overline{q}_3)^c (q_2)^{3}
\rangle^3$ and $| (q_2 \overline{q}_3)^c (q_1)^{3} \rangle^3$
where now $c = 1$ or $8$ (we omit the lower spin index in this
computation). By direct calculation, one obtains the colour
crossing:
\begin{eqnarray}\label{6}
    V_c \equiv
    \left(
\begin{array}{c}
| (q_1 q_2)^6 (\overline{q}_3)^{\overline{q}_3} \rangle^3 \\
| (q_1 q_2)^{\overline{3}} (\overline{3})^{\overline{q}_3}
\rangle^3
\end{array}
     \right) & = &
     \left[
     \begin{array}{cc}
     \sqrt{\frac{2}{3}} & \sqrt{\frac{1}{3}} \\
      - \sqrt{\frac{1}{3}} &  \sqrt{\frac{2}{3}}
     \end{array}
     \right] \ \
     \left(
\begin{array}{c}
| (q_2 \overline{q}_3)^1 (q_1)^{3} \rangle^3 \\
| (q_2 \overline{q}_3)^8 (q_1)^{3} \rangle^3
\end{array}
     \right) \nonumber \\
      & = &
      \left[
     \begin{array}{cc}
     \sqrt{\frac{2}{3}} & \sqrt{\frac{1}{3}} \\
      \sqrt{\frac{1}{3}} &  - \sqrt{\frac{2}{3}}
     \end{array}
     \right] \ \
     \left(
\begin{array}{c}
| (q_1 \overline{q}_3)^1 (q_2)^{3} \rangle^3 \\
| (q_1 \overline{q}_3)^8 (q_2)^{3} \rangle^3
\end{array}
     \right)
\end{eqnarray}
from where we can also derive the (inverse) expressions of $| (q_1
\overline{q}_3)^c (q_2)^{3} \rangle^3$ and $| (q_2
\overline{q}_3)^c (q_1)^{3} \rangle^3$ in terms of $| (q_1 q_2)^c
(\overline{q}_3)^{\overline{3}} \rangle^3$. It is then
straightforward to derive the $H_C$ matrix.

A similar technic will allow to construct the $2 \times 2 \ H_S$
matrix, and we finally give the complete expression for the
colour magnetic Hamiltonian $H_{CM}$ acting on the 4-dim vector
$\overrightarrow{\phi} = (\phi_1, \phi_2, \phi_3, \phi_4)$:\\
$H_{CM}$ =
\begin{equation}\label{7}
 - \left[
\begin {array}{cccc}
\frac{4}{3}\,{\it C_{12}}+\frac{20}{3} ({\it C_{13}} +{\it
C_{23}}) & 4 \sqrt {2} \left( {\it C_{13}}-{\it C_{23}} \right) &
\frac{10}{\sqrt {3}} \left( {\it C_{13}}-{\it C_{23}} \right)
&2\sqrt {6}
  \left( {\it C_{13}}+{\it C_{23}} \right) \\\noalign{\medskip}
  4 \sqrt {2} \left( {\it C_{13}}-{\it C_{23}} \right) & -\frac{8}{3}\,{\it C_{12}}+\frac{8}{3}\,({
\it C_{13}} + {\it C_{23}})& 2 \sqrt{6} \left( {\it C_{13}}+{\it
C_{23}} \right) & \frac{4}{\sqrt{3}} \left( {\it C_{13}}-{\it
C_{23}} \right) \\\noalign{\medskip} \frac{10}{\sqrt {3}} \left(
{\it C_{13}}-{\it C_{23}}  \right) &2\sqrt {6} \left( {\it
C_{13}}+{\it C_{23}} \right) &-4\,{\it
C_{12}}&0\\\noalign{\medskip} 2\, \sqrt{6} \left( {\it
C_{13}}+{\it C_{23}} \right) & \frac{4}{\sqrt {3}} \left( {\it
C_{13}}-{\it C_{23}} \right) &0&8\,{ \it C_{12}}
\end {array}
\right]
\end{equation}
\vspace{4mm}

It is easily seen from this matrix that, if we impose flavour
symmetry for the two quarks ($C_{12}$ = $C_{23}$), we get a matrix
operating over two invariant subspaces $\{ \phi_1 , \phi_4 \}$ and $\{ \phi_2 , \phi_3 \}$ respectively.\\
If,in addition we impose full flavour symmetry for the
  interaction and assume that the qq and $q\bar{q}$ interactions are the same
  (so that $C_{ij}=C$) we have the matrix\\
\begin{equation}\label{8}
H_{\mathrm{CM}} = -C \times \left[ \begin {array}{cccc} {\frac
{44}{3}}&0&0&4\,\sqrt {6}
\\\noalign{\medskip}0& \frac{8}{3} &4\,\sqrt {6}&0\\\noalign{\medskip}0&4\,
\sqrt {6}&-4&0\\\noalign{\medskip}4\,\sqrt {6}&0&0&8\end {array}
  \right]
\end{equation}

\vspace{4mm}

\noindent and we fall back on the old results 
\cite{Chan:1978nk} ,\cite{Mulders:1978cp}  where the
eigenvalues of the colourmagnetic interaction are -21.88C and
-0.98C for the case when the two quarks are coupled symmetrically
  in colour-spin .For antisymmetric colour spin the eigenvalues are
  -9.68C and 11.02C.

  In no case are the eigenvectors corresponding to sharp
  values of the total colour-spin. They are mixtures of the 6
  and 120 dimensional representations as well as of the 6 and $\overline{84}$
  representations of the colour spin $SU(6)_{cs}$ algebra when
  considering the $(qq\overline{q})^3_{1/2}$ system. Indeed,
  performing the product of $SU(6)$ representations:
  \begin{equation}\label{9}
    21 \times \overline{6} = 6 + 120 \ \ \ \ \mbox{and} \ \ \ \ 15
    \times \overline{6} = 6 + \overline{84}
\end{equation}
and examining the corresponding $SU(3) \times SU(2)$
decompositions
\begin{equation}\label{10}
    6 = (3 \ , \frac{1}{2}) \ \ \ \ \ \ \ \  120= (3+15 \ ,  \frac{1}{2} +
    \frac{3}{2}) + (\overline{6} \ , \frac{1}{2}) \ \ \ \ \ \ \ \ \overline{84}
    = (15 \ , \frac{1}{2}) + (3+\overline{6} \ , \frac{1}{2} +
    \frac{3}{2})
\end{equation}
one easily remarks that both the 6 and 120 $SU(6)$ representations
contain a triplet of colour and doublet of spin, and that is also
the case for the couple of representations 6 and $\overline{84}$.

  Moreover, if we decouple the antiquark (going to the heavy quark limit or
  considering
  relative spatial wave functions that have no $s$-wave overlap)
  putting $C_{13}$ = $C_{23}=0$, the effective Hamiltonian $H_{CM}$ is
  diagonal, with elements
  which are the well known colour magnetic energies for colour  sextet and triplet diquarks.

  As has been remarked before, if the two quarks are identical in
  flavour, the matrix is $2 \times 2$ and the states $\phi_1$ and
  $\phi_4$ disappear.

  After the invention of flavour symmetry groups, it has become the
  custom to mark flavour combinations in multiplets of the flavour
  symmetry groups in accordance with the {\em generalized} Pauli
  principle. In the flavour symmetry limit, the states $\phi_1$ and
  $\phi_4$ which have the two quarks  in the symmetric colour spin
  representation 21 are associated with the flavour $SU(3)$
  representation $F = \overline{3}$, while the states $\phi_2$ and
  $\phi_3$ stand in the $F=6$ representation as the two quarks are
  in the antisymmetric representation of colour spin.

  Note that the flavour content $(qq\overline{q})$ is
  $\overline{3}\times \overline{3} = 3 + \overline{6}$ for $\phi_1$ and
  $\phi_4$ and $6 \times \overline{3} = 3 + 15$ for $\phi_2$ and
  $\phi_3$.

  When the "triquark" $(qq\overline{q})$ is combined with the
  (most strongly bound) "diquark" $(qq)$ which has
  $c=\overline{3}, s=0$ and flavour $F=\overline{3}$, the total
  $(qqqq\overline{q})$ states containing $\phi_1$ and
  $\phi_4$ will be in the flavour representation $(3+\overline{6}
  ) \times \overline{3} = 1+8+8+\overline{10}$, while the states
  containing $\phi_2$ and $\phi_3$ will be in the  $(3+15) \times
  \overline{3}= 1+8+8+10+27$ flavour representations.

  The representations $\overline{10}$ in the first group, and 27 in
  the second group, manifestly contain exotics.

  As we have seen, $\phi_1$ and
  $\phi_4$ will mix as well as $\phi_2$ and
  $\phi_3$ if there is colour magnetic interaction
  $(C_{q\overline{q}} \neq 0)$ between the antiquark and the
  quarks. When flavour symmetry is broken, all states will in
  general mix: this corresponds to mixing of states in different
  "flavour" representations.

  If we use isospin symmetric $u$ and
  $d$ quarks, then $C_{13} = C_{23}$ and states with different
  flavour symmetry will not mix. This is the case for all models
  of the exotic $\Theta^+$ which is assumed to be $(ud \ ud\overline{s})$,
  and would therefore only belong to the $F=\overline{10}$
  representation.

  On the contrary, for $\Xi^{--}$ which is of the form $(ds \ ds \
  \overline{u})$, the colour magnetic interaction between
  $(d\overline{u})$ and $^(s \overline{u})$, with $(C_{13} \neq
  C_{23})$, will mix the $(ds \ \overline{u})$ in
  the$F=\overline{6}$ and in the $F=15$ representations. Therefore
  the exotic $(ds \ ds \ \overline{u})$ appears in both $F=\overline{10}$ and $F=27$
  representations and these will mix.

\indent

We next apply
these results on the $qq \bar{q}$ cluster with broken flavour symmetry to
some current models of pentaquark states.  As we shall see, the breaking of
flavour symmetry has considerable effects on the mass spectrum.
 As already remarked, the coefficients
$C_{ij}$ depend on the relative s-wave
overlap of the spatial wave functions of the quarks $q_i$ and $q_j$.\\
 Considering, in a first  picture, the case of a  "triquark"  bound to a "diquark"
 in  a relative p-wave \cite{kl1}, we do what is usually
  done: We take values for the
coefficients that are close to the ones which are quite successful
when applied
to ground state baryons and mesons.\\
We use  $C_{ij}$=20MeV for the interaction between nonstrange
quarks and antiquarks, $C_{ij}=12.5$ MeV between one strange and one
nonstrange quark (antiquark),$C_{ij}$=5MeV between one nonstrange
quark and a charm quark (antiquark). It is then straightforward to
calculate the binding of the "triquark" coming from
$H_{\mathrm{CM}}$.\\
The lowest eigenvalue of $H_{\mathrm{CM}}$ for $(ud\bar{s})$ is then -300Mev, for
for $(ds\bar{u})$ it is
-357MeV and for  $(ud\bar{c})$ the lowest eigenvalue for $H_{\mathrm{CM}}$ is -186MeV. These values are
somewhat different from the values one gets when
the mixing of
states $\phi_i$ is ignored \cite{kl1} .

The triquarks above are combined with the lowest mass "diquark"
carrying colour $\bar{\mathbf{3}}$ and spin 0 in a relative
p-state.(This makes it not too unreasonable to neglect antisymmetrization
 between identical quarks in different clusters.). The mass defect due to the colourmagnetic interaction for
this diquark cluster ($q_i$ $q_j$)  is 8$C_{ij}$. If one assumes that
the cost of a p-wave excitation is somewhat similar in all cases,  one
 gets relations between masses M($qqqq\bar{q}$ ) of different exotic states that
 fall in the same "flavour multiplet":
 \begin{equation}\label{11}
 M(sdsd\bar{u} )-M(udud\bar{s} )=   m_s -m_u  -3MeV.
\end{equation}
\begin{equation}\label{12}
 M(udud\bar{c} )-M(udud\bar{s} )=  m_c -m_s   +114 MeV.
\end{equation}

Here $m_i$ denotes the effective mass of each quark:
\begin{equation}\label{12bis}
 m_d \approx     m_u \approx  360MeV,\ m_s \approx  540MeV, \ m_c \approx   1710MeV.
\end{equation}

  It is difficult to be encouraged by these results. The absolute value of
 M($udud\bar{s}$ ) would be $\approx$   1520MeV without the cost of the L=1
 excitation.
 To use the coefficients  $C_{ij}$ as free parameters is possible
 but it is not an attractive method. \\

It is problematic to get states in multiquark
spectroscopy through the colour magnetic interaction that are
light enough to accommodate  the $ \Theta^+ (1540)$. We have
computed the eigenvalues of many other types of quark clusters
with flavour symmetry breaking colour magnetic interactions \cite{HogSorb04} and no
one gives states with masses smaller than the
$(qq\overline{q})^3_{1/2} - (qq)^{\overline{3}}$
.

 If we normalize M($udud\bar{s}$ ) to 1540MeV, we could "predict" $ \Xi {^{--}}$
 at 1720MeV, far from 1862 MeV, and the Hera state $(udud\bar{c})$  at $\approx $
 2825MeV (whereas the mass observed \cite{H1} is $\approx $  3100MeV).\\
 Moreover, from the colour crossing matrix we see that  all  $(qq\bar{q})$ s-wave
 clusters that carry colour ${\mathbf{3}}$ contain $(q\bar{q})$ systems that are
 colour singlets. In old days, that was supposed to lead to "superallowed decays"
 and ions(clusters) carrying colour ${\mathbf{3}}$ were ignored when one tried to
 predict narrow multiquark states. \\
 In our original study \cite{Hogaasen:1978jw} of $(4q\overline{q})$ states
considered as two colour non singlet clusters separated by a
relative angular momentum,  we omitted the colour triplet configurations,
 as the $(qq\overline{q}) -(qq)$ one considered in ref(10)
 and in this letter, as well as the $(4q)-(\overline{q})$
and $(3q\overline{q}) - ( q)$ ones, due to the presence of a
colour singlet $(q\overline{q})$ part in one of the clusters.
The $(q\overline{q})$
singlet cluster is explicit in eq.(\ref{6})).\\

 Let us now turn to another picture\cite{jw1}  .Here two diquarks carrying colour
 $\bar{\mathbf{3}}$ and spin zero is in a
 relative p-wave and the antiquark is in a s-wave relative to
  the overall center of mass.. One could believe that
 our considerations have no relevance for this picture. Indeed, if none of the
 quarks in their cluster were in a relative s-wave with the antiquark, that would
 be so. If this is not the case, the presence of the antiquark will influence
 the state so that the diquark  $\bar{\mathbf{3}}$ and spin zero will mix with
 (generally) the diquarks carrying colour $\bar{\mathbf{3}}$  spin1,colour
 $\bar{\mathbf{6}}$ spin zero and colour $\bar{\mathbf{3}}$ spin1.As explained before,
  for $\Theta^+(1540)$ the mixing would be between only two states.
 In a recent
 attempt to
 dynamically generate a "diquark" "diquark"
 antiquark description \cite{dudek}  , there certainly is a s-wave overlap between the antiquark
 and the quarks.To understand quantitatively the consequences of this mixing would require
 a full calculation for the $qqqq\bar{q}$. We strongly encourage such studies as
 the ones of \cite{dudek} and \cite{jennings} with the application of
  our formula (7).We have noted that our results for the lowest eigenvalues of $H_{\mathrm{CM}}$
  for the $udud\bar{s}$ configuration, relevant for calculations for $\Theta^+(1540)$,
   fall between what is calculated for the zero range
   and the finite range version of the colour magnetic spatial dependence
    in \cite{jennings}.

We now return to the configuration $(qq\overline{q}) - (qq)$ of
two clusters with colour $3-\overline{3}$ and relative angular
momentum $L=1$,and note that it does not seem adapted for understanding the
experimental $\Xi^{--} - \Theta^+$ mass difference.

As remarked,the easy way out to lower the "predicted" masses would be to use the coefficients $C_{ij}$'s as free
parameters. The problem remains however if one wants to put states
as $\Xi^{--}$ and $\Theta^+$ in the same  (mixed) flavour
multiplet. Let us consider the $(qq\overline{q})^3_{1/2}$ cluster as
an example. We take $C_{ud} = C$ as a free parameter and keep the
ratio $\frac{C_{us}}{C_{ud}} = \frac{5}{8}$ as before.
(This seems to us to be quite a reasonable approximation). Then, one
would get a "mass" for the ($ud\overline{s})$ cluster:
\begin{equation}\label{13}
    M(ud\overline{s}) = 2m_u + m_s - 15.01 C +E_L
\end{equation}
and for the $(ds\overline{u})$ cluster
\begin{equation}\label{14}
M(ds\overline{u}) = 2m_u + m_s - 17,85 C + E_L
\end{equation}
where $E_L$ is the $L=1$ excitation energy.

So if we decrease the masses by increasing $C$, the mass
difference $M(ds\overline{u}) - M(ud\overline{s}) = - 2,84 C$ will
also decrease.As we have seen,the mass difference was too small to start with.
The situation will be worse.\\

In the $(qq)(qq)\overline{q}$ configuration  \cite{jw1} , one does not have
this difficulty. Here an increase in $C$ will lower the
theoretical masses and at the same time increase the mass
difference of the $(ds)(ds)\overline{u}$ and
$(ud)(ud)\overline{s}$ configuration. A modest increase of
$C_{ud}$ from 20 MeV to 24 MeV will give the observed mass
difference between $\Xi^{--}$ and $\Theta^+$ while in the same
time the theoretical mass of $\Theta^+$ will decrease by 65 MeV.
 But again:To understand how the $\Theta^+$ can be at
such a low mass as 1540MeV is difficult.
Some interesting attempts based on different approaches are proposed
\cite{stancu,closedud,carlson,bucsorba,kochelev}
concerning this question as well as the narrowness of the observed states.\\

Acknowledgment: It is a pleasure to thank J.M.Richard for a useful
discussion.

\end{document}